\documentclass[letter,twocolumn]{jpsj3} 
\usepackage{color}
\title{
Magnetization Process 
of the Spin-$S$ Kagome-Lattice Heisenberg Antiferromagnet 
}
\catcode`\@=11
\def\simle{\mathrel{\mathpalette\@versim<}}   
\def\simge{\mathrel{\mathpalette\@versim>}}   
\def\@versim#1#2{\lower2.5pt\vbox{\baselineskip0pt \lineskip-.5pt
   \ialign{$\m@th#1\hfil##\hfil$\crcr#2\crcr\sim\crcr}}}
\catcode`\@=12

\author{Hiroki Nakano$^{1}$
\thanks{E-mail: hnakano@sci.u-hyogo.ac.jp} 
and 
T\^oru Sakai$^{1,2}$
\thanks{E-mail: sakai@spring8.or.jp}
}

\inst{$^{1}$Graduate School of Material Science, 
University of Hyogo,
Kamigori, 
Hyogo 678-1297, Japan \\
$^{2}$
Japan Atomic Energy Agency, SPring-8, 
Sayo, Hyogo 679-5148, Japan 
}

\recdate{\today}

\abst{
The magnetization process of the spin-$S$ Heisenberg 
antiferromagnet on the kagome lattice is 
studied by the numerical-diagonalization method. 
Our numerical-diagonalization data for small finite-size clusters 
with $S=1$, 3/2, 2, and 5/2 suggest 
that a magnetization plateau appears 
at one-third of the height of the saturation 
in the magnetization process irrespective of $S$. 
We discuss the $S$ dependences of the edge fields and the width 
of the plateau in comparison with recent results obtained 
by real-space perturbation theory. 
}

\begin{document}
\maketitle


Frustrated spin systems have attracted much attention 
from many condensed-matter physicists. 
One of the fascinating systems among them is 
the kagome-lattice Heisenberg antiferromagnet. 
Unfortunately, 
our understanding of this system is still far from complete 
in spite of many experimental and theoretical studies.  
In the $S=1/2$ system, in particular, 
discoveries of several realistic materials such as  
herbertsmithite\cite{Shores_herbertsmithite2005,Mendels_herbertsmithite2010}, 
volborthite\cite{Yoshida_jpsj_volborthite2009,Yoshida_prl_volborthite2009}, 
and 
vesignieite\cite{Okamoto_jpsj_vesignieite2009,Okamoto_prb_vesignieite2011} 
have accelerated theoretical studies\cite{Lecheminant,Waldtmann,Hida_kagome2,
cabra2002,Honecker0,Honecker1,cabra2005,Cepas,Spin_gap_Sindzingre,
kgm_ramp,Sakai_HN_PRBR,kgm_gap,Honecker2011,capponi2013,Jiang2008,
Yan_Huse_White_DMRG,Depenbrock2012,SNishimoto_NShibata_CHotta,
HN_Sakai_PSS,HN_Sakai_SCES,HN_TSakai_kgm_1_3,Iqbal2014,Iqbal2015}. 
However, there remain some unresolved issues; 
one of them is the spin-gap problem 
of whether the spin excitation above the singlet ground state 
is gapped or gapless. 

On the other hand, fewer studies on $S>1/2$ cases have been 
carried out. 
As candidate $S=1$ kagome-lattice systems, 
$m$-MPYNM$\cdot$BF$_{4}$\cite{NWada_S1kgm1997,TMatsushita_S1kgm1997}, 
NaV$_3$(OH)$_6$(SO$_4$)$_2$\cite{JNBehera_S1kgm2002}, 
[C$_6$N$_2$H$_8$][NH$_4$]$_2$[Ni$_3$F$_6$(SO$_4$)$_2$]\cite{JNBehera_S1kgm2006}, 
and 
KV$_{3}$Ge$_{2}$O$_{9}$\cite{SHara_HSato_YNarumi_S1kgm2012} are known. 
Theoretical studies\cite{HAsakawa,Hida_kagome1,Hida_kagome2,Li_S1kgm,
SNishimoto_MNakamura_condmat1409-5870-s1kgm,
Liu_PRB91_060403} for the $S=1$ case 
are also limited. 
Studies on the $S>1$ cases have only started recently; 
candidate kagome-lattice systems of 
Cs$_{2}$Mn$_{3}$LiF$_{12}$\cite{KKatayama_NKurita_HTanaka_JPS_S2kgm} 
for $S=2$ 
and 
NaBa$_{2}$Mn$_{3}$F$_{11}$\cite{KNawa_JPS_S2_5kgm} for $S=5/2$ 
have been reported, 
together with theoretical studies\cite{Bergman_JPCM19_145204,
Gotze_PRB2011,Zhitomirsky} 
as well as an analysis based on the semiclassical limit\cite{cabra2005}. 

Under these circumstances, then, we are faced with a question: 
do any systematic behaviors exist 
in the spin-$S$ kagome-lattice Heisenberg antiferromagnet 
under magnetic fields? 
The purpose of this letter is to extract 
such systematic behavior of the magnetization processes 
of this model for various $S$ by numerical-diagonalization calculations 
that are unbiased against approximations. 
With the same motivation, Zhitomirsky recently investigated 
the frustrated Heisenberg model under magnetic fields 
by real-space perturbation theory taking into account fluctuations 
around a classical configuration\cite{Zhitomirsky}. 
He found that 
in the magnetization process of the kagome-lattice antiferromagnet, 
the so-called $uud$ state is stable 
at one-third of the height of the saturation, 
at which a magnetization plateau appears 
irrespective of the value of $S$. 
He also derived an expression for the $1/S$ expansion 
for both the edge fields of this height. 
The comparison between the present numerical-diagonalization results 
and the results from real-space perturbation theory 
should contribute to our understanding of the frustration effect 
in the kagome-lattice antiferromagnet. 


The Hamiltonian that we study in this research is given by 
${\cal H}={\cal H}_0 + {\cal H}_{\rm Zeeman}$, where 
\begin{equation}
{\cal H}_0 = \sum_{\langle i,j\rangle} J 
\mbox{\boldmath $S$}_{i}\cdot\mbox{\boldmath $S$}_{j} , 
\label{H_undistorted_kagome}
\end{equation}
and 
\begin{equation}
{\cal H}_{\rm Zeeman} = - H \sum_{j} S_{j}^{z} .  
\label{H_zeeman}
\end{equation}
Here, $\mbox{\boldmath $S$}_{i}$ 
denotes the spin operator 
at site $i$, where the sites are the vertices of the kagome lattice. 
The spin operator satisfies $\mbox{\boldmath $S$}_{i}^2=S(S+1)$. 
The sum of ${\cal H}_0$ runs over all the nearest-neighbor pairs 
in the kagome lattice. 
Energies are measured in units of $J$; 
hereafter, we set $J=1$.  
The number of spin sites is denoted by $N_{\rm s}$, 
where $N_{\rm s}/3$ is an integer. 
We impose the periodic boundary condition 
for clusters with site $N_{\rm s}$. 

We calculate the lowest energy of ${\cal H}_0$ 
in the subspace belonging to $\sum _j S_j^z=M$ 
by numerical diagonalizations 
based on the Lanczos algorithm and/or the Householder algorithm. 
The energy is denoted by $E(N_{\rm s},M)$, 
where $M$ takes an integer or a half odd integer value 
up to the saturation value $M_{\rm s}$ ($=S N_{\rm s}$). 
We often use the normalized magnetization $m=M/M_{\rm s}$. 
Part of the Lanczos diagonalizations were carried out 
using the MPI-parallelized code, which was originally 
developed in the study of Haldane gaps\cite{HN_Terai}. 
The usefulness of our program was previously confirmed in large-scale 
parallelized calculations\cite{kgm_gap,s1tri_LRO,HN_TSakai_kgm_1_3}. 

The magnetization process for a finite-size system is obtained 
by considering the magnetization increase from $M$ to $M+1$ in the field 
\begin{equation}
H=E(N_{\rm s},M+1)-E(N_{\rm s},M),
\label{field_at_M}
\end{equation}
under the condition that the lowest-energy state 
with magnetization $M$ and that with magnetization $M+1$ 
become the ground state in specific magnetic fields.  


\begin{figure*}[tb]
\begin{center}
\includegraphics[width=15cm]{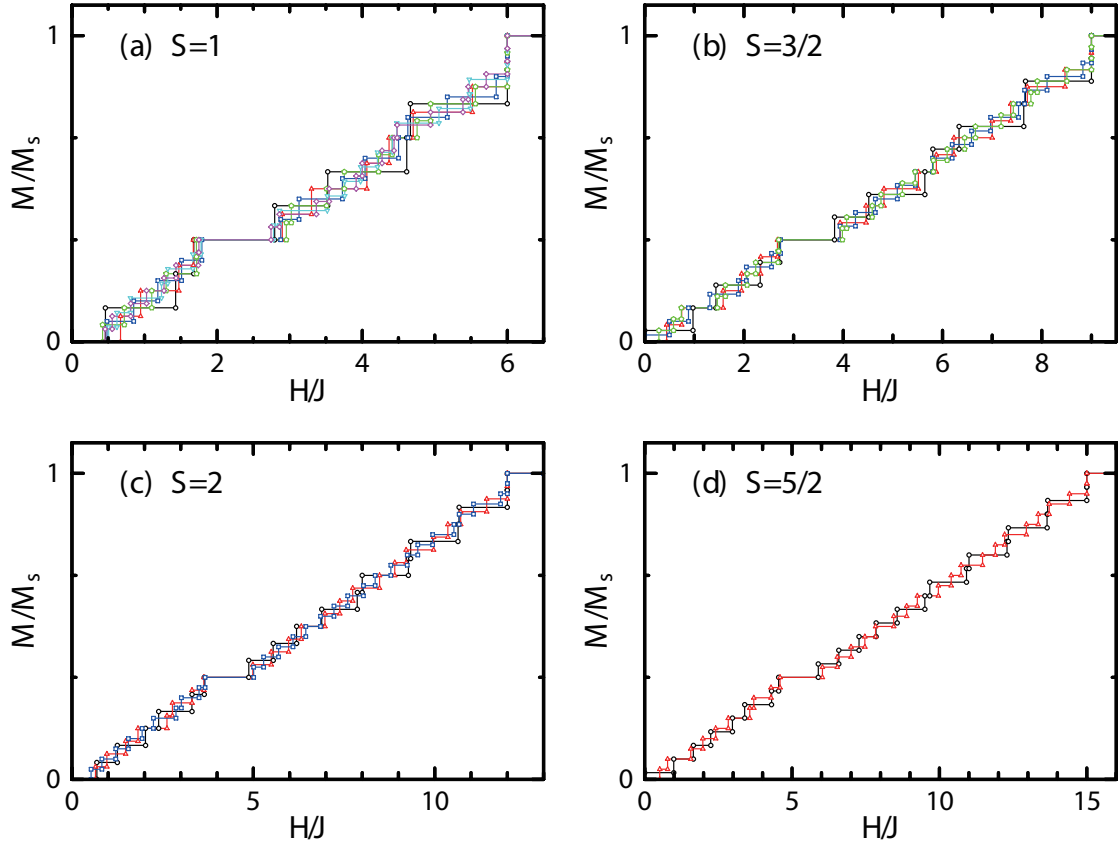}
\end{center}
\caption{(Color) 
Magnetization processes for $S=1$, 3/2, 2, and 5/2 
in (a), (b), (c), and (d), in which 
the maximum sizes are $N_{\rm s}=24$, 18, 15, and 12, respectively. 
Black circles, red triangles, blue squares, green pentagons, 
light-blue inverted triangles, and violet diamonds 
linked by lines of the same color denote 
the cases of $N_{\rm s}=9$, 12, 15, 18, 21, and 24, respectively. 
}
\label{fig1}
\end{figure*}
First, let us present our results of the magnetization processes 
for $S=1$, 3/2, 2, and 5/2; results are shown in Fig.~\ref{fig1}. 
The maximum sizes of the clusters treated in this study 
are $N_{\rm s}=24$, 18, 15, and 12 for $S=1$, 3/2, 2, and 5/2, 
respectively. 
The cluster shapes we calculated are the same as those 
in Ref.~\ref{Hida_kagome2}. 
Note here that, the shapes for $N_{\rm s}=9$, 12, and 21 
are rhombic and that the others are nonrhombic. 
In Fig.~\ref{fig1}(a), 
the entire range of the cases up to $N_{\rm s}=18$ 
and part of the range in the case of $N_{\rm s}=21$ were already 
reported in Ref.~\ref{Hida_kagome2} and 
the rest of the range in the case of $N_{\rm s}=21$ and 
the results for $N_{\rm s}=24$ 
are additionally presented in the present study. 
For the cases of $S>1$, 
there are no reports on numerical-diagonalization calculations 
of the magnetization processes 
to the best of our knowledge. 
The most noteworthy behavior is observed 
at one-third of the height of the saturation, 
where behavior similar to a magnetization plateau appears 
irrespective of the value of $S$.  
A detailed discussion concerning the edges and the width 
of this height will be given later. 
The next characteristic behavior is 
a jump near the saturation. 
Note here that 
all the states within the jump are numerically degenerate 
at the saturation field $H_{\rm s} (=6JS)$. 
A similar jump is known to occur in several 
cases\cite{Schnack_jump,Honecker0,HJSchmidt_jump,Zhitomirsky_Tsunetsugu_jump}. 
Regarding the existence or the absence of the degeneracy, 
this behavior is different from 
the magnetization jump observed in square-kagome-lattice 
and Cairo-pentagon-lattice antiferromagnets, and 
so on\cite{shuriken_lett,HN_kgm_dist,HNakano_Cairo_lt,Isoda_Cairo_full}. 
Although 
this behavior occurs irrespective of the value of $S$, 
the skip of $m$ at the jump gradually decreases as $S$ is increased. 
Around $m=(9S-2)/(9S)$ near the jump for $S=1$, 3/2, 2, and 5/2, 
a region where the gradient of the magnetization process seems small 
may exist, 
although the behavior is very faint. 
To clarify the existence of the magnetization plateau 
at this height, future studies of larger systems are required. 
Note here that 
this height does not correspond 
to the $m=7/9$ plateau near the jump in the $S=1/2$ case 
in Ref.~\ref{Zhitomirsky_Tsunetsugu_jump} 
but that it corresponds to $m=5/9$ for $S=1/2$, 
at which the existence of another plateau has been pointed out 
for the $S=1/2$ kagome-lattice 
antiferromagnet\cite{SNishimoto_NShibata_CHotta,capponi2013}. 
There appears an overhanging behavior at $m=17/21$ 
only for $N_{\rm s}=21$ in the $S=1$ case; 
this may be an artifact owing to a finite-size effect  
beacuse it is very small and beacuse it does not appear for $N_{\rm s}=24$. 

Hereafter, we focus our attention on the behavior at $m=1/3$. 

\begin{figure}[tb]
\begin{center}
\includegraphics[width=7cm]{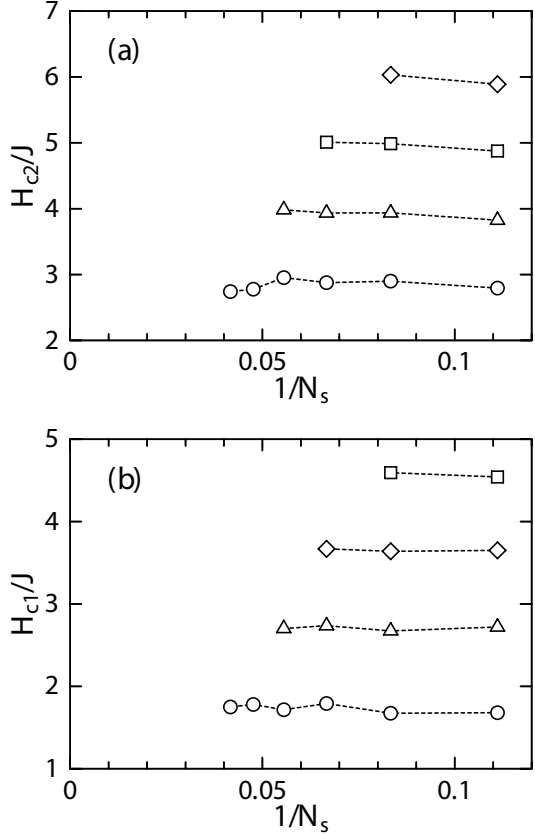}
\end{center}
\caption{
System-size dependence of (a) $H_{\rm c2}$ and (b) $H_{\rm c1}$. 
Circles, triangles, squares, and diamonds denote 
the cases of $S=1$, 3/2, 2, and 5/2, respectively.  
}
\label{fig2}
\end{figure}
Next, let us examine the system-size dependence 
of the edges at $m=1/3$, i.e., the lower-field edge $H_{\rm c1}$ 
and the higher-field edge $H_{\rm c2}$; 
results are shown in Fig.~\ref{fig2}. 
In all cases of $S$, the size dependences of $H_{\rm c1}$ and $H_{\rm c2}$ 
are not large. 
Recall here that in the case of $S=1/2$, 
the discontinuous size dependence between $N_{\rm s}=18$ and 21 
is known to occur in $H_{\rm c2}$\cite{Hida_kagome2}; 
the decrease is about 20\%. 
In the case of $S=1$, on the other hand, 
the decrease between $N_{\rm s}=18$ and 21 is about 6\%, 
which is much smaller than that in the case of $S=1/2$. 
It is therefore reasonable 
to use the values of $H_{\rm c1}$ and $H_{\rm c2}$ 
of the largest cluster for each $S$ 
as substitutes for those of the infinite-size system 
without extrapolation 
when we carry out a detailed analysis. 
Note also that 
the small size dependences of $H_{\rm c1}$ and $H_{\rm c2}$ 
are related to the properties of the $m=1/3$ states. 
After the $m=1/3$ states were studied in an analysis 
based on the semiclassical limit\cite{cabra2005}, 
it was pointed out that the $m=1/3$ states reveal a nine-site 
structure in the unit cells of the spin states 
from the analysis of an effective Hamiltonian obtained 
by perturbation theory from the Ising limit\cite{Bergman_JPCM19_145204},  
although Refs.~\ref{cabra2002} and \ref{Bergman_JPCM19_145204} 
did not clarify the existence of the plateau 
or present estimates of $H_{\rm c1}$ and $H_{\rm c2}$. 
If the states form the nine-site structure, 
the states are energetically stable when $N_{\rm s}/9$ is an integer; 
on the other hand, the energies are higher 
when $N_{\rm s}/9$ is not an integer 
than when $N_{\rm s}/9$ is an integer. 
This arguement suggests that for finite-size $H_{\rm c1}$ and $H_{\rm c2}$, 
$H_{\rm c1}$ ($H_{\rm c2}$) 
becomes lower (higher) only when $N_{\rm s}/9$ is an integer. 
However, such behavior is not observed in Fig.~\ref{fig2}.  
Thus, the present results do not support 
the nine-site structure in the $m=1/3$ states. 
The same situation was pointed out in Ref.~\ref{HN_TSakai_kgm_1_3} 
for the $S=1/2$ kagome-lattice antiferromagnet. 
For future studies, numerical data for $H_{\rm c1}$ and $H_{\rm c2}$ 
are shown in Table~\ref{table1} 
together with the singlet ground-state energy $E(N_{\rm s},0)$. 
\begin{table}[hbt]
\caption{
Edge fields for the $m=1/3$ height in the magnetization process 
of the spin-$S$ kagome-lattice Heisenberg antiferromagnet 
for the largest cluster treated in the present study. 
We also present the energy per site of the singlet ground 
state\cite{comment_GS}. 
}
\label{table1}
\begin{center}
\begin{tabular}{ccccc}
\hline
 $S$ & $N_{\rm s}$ & $H_{\rm c1}/J$ & $H_{\rm c2}/J$ & 
$ E(N_{\rm s},0)/(N_{\rm s}JS^2)$\\
\hline
 1   & 24 & 1.7502 & 2.7430 & -1.4266894\\
 3/2 & 18 & 2.7016 & 3.9827 & -1.2895265\\
 2   & 15 & 3.6697 & 5.0085 & -1.2259126\\
 5/2 & 12 & 4.5944 & 6.0308 & -1.1835511\\
\hline
\end{tabular}
\end{center}
\end{table}

\begin{figure}[tb]
\begin{center}
\includegraphics[width=7cm]{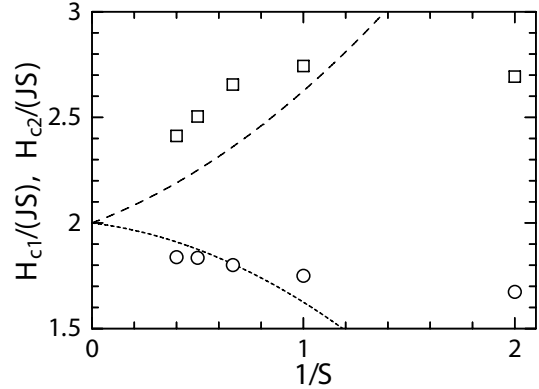}
\end{center}
\caption{
$S$-dependence of $H_{\rm c1}$ and $H_{\rm c2}$ of the largest-size clusters. 
Circles and squares represent the results 
of $H_{\rm c1}$ and $H_{\rm c2}$ for the largest-size clusters, respectively. 
Equations (\ref{Zhitomirsky_Hc1}) and (\ref{Zhitomirsky_Hc2}) 
are drawn as dashed and broken curves, respectively. 
}
\label{fig3}
\end{figure}
Next, let us examine the $S$ dependence of $H_{\rm c1}$ and $H_{\rm c2}$. 
We plotted $H_{\rm c1}$ and $H_{\rm c2}$ for the largest cluster 
as a function of $1/S$; the result is shown in Fig.~\ref{fig3} 
together with $H_{\rm c1}$ and $H_{\rm c2}$ for the $N_{\rm s}=42$ cluster 
in the $S=1/2$ case reported in Ref.~\ref{HN_TSakai_kgm_1_3}  
as the largest-cluster result for $S=1/2$. 
We also draw the curves of the expressions
\begin{eqnarray}
\label{Zhitomirsky_Hc1}
\frac{H_{\rm c1}}{JS}&=&2-\frac{1}{8S} - \frac{1}{4S^2} \\
\label{Zhitomirsky_Hc2}
\frac{H_{\rm c2}}{JS}&=&2+\frac{3}{8S} + \frac{1}{4S^2},  
\end{eqnarray}
derived in Ref.~\ref{Zhitomirsky}. 
One observes that the dependence of $H_{\rm c2}$ changes 
between $S=1$ and $S=3/2$ 
and 
that the numerically obtained $H_{\rm c2}$ for $S \ge 3/2$ 
approaches a value of 2 with increasing $S$, 
which is the value for an infinite $S$, namely, the classical case. 
The agreement of $\lim_{S\rightarrow\infty} H_{\rm c2}$ with the classical value 
suggests that the substitution explained in the above is reasonable. 
Note here that the $1/S$ dependence of the numerically obtained $H_{\rm c2}$ 
is convex upward. 
This convex dependence is clearly different from Eq.~(\ref{Zhitomirsky_Hc2}). 
On the other hand, $H_{\rm c1}$ shows an almost linear dependence on $1/S$; 
the line seems to approach a value that is slightly smaller than 2, 
which is also the value for the classical case. 
The reason for the difference of this value from the classical value 
is unclear at present. 
The dependence of $H_{\rm c1}$ may change above $S=5/2$ 
if we assume a continuous dependence toward the classical limit.  
Thus, the $1/S$ dependence must be concave upward when $S$ is large.  
This concave dependence is clearly different from Eq.~(\ref{Zhitomirsky_Hc1}). 
Note additionally that 
the almost smooth $S$-dependences of our results in Fig.~\ref{fig3} 
do not support the existence of the nine-site structure in the $m=1/3$ state 
pointed out in Ref.~\ref{Bergman_JPCM19_145204}.

\begin{figure}[tb]
\begin{center}
\includegraphics[width=7cm]{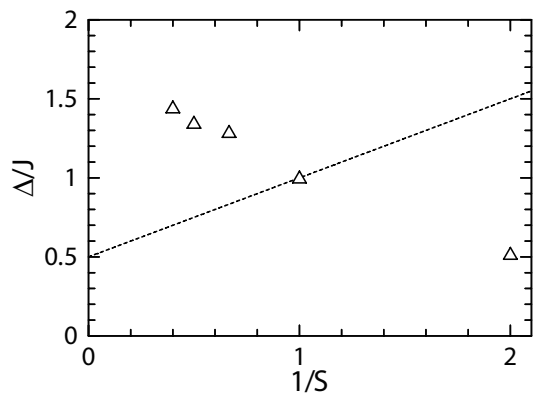}
\end{center}
\caption{
Width $\Delta$ for the largest-size clusters plotted 
as a function of $1/S$. 
The dashed line denotes Eq. (\ref{Zhitomirsky_delta}). 
}
\label{fig4}
\end{figure}
Finally, let us analyze the width of the height at $m=1/3$ 
in the magnetization process, 
where the width is given by $\Delta= H_{\rm c2} - H_{\rm c1}$; 
the result is shown in Fig.~\ref{fig4}, 
in which we also draw the line 
\begin{equation}
\frac{\Delta}{J}=\frac{1}{2} + \frac{1}{2S}, 
\label{Zhitomirsky_delta}
\end{equation}
obtained from Eqs.~(\ref{Zhitomirsky_Hc1}) and (\ref{Zhitomirsky_Hc2}). 
The most striking feature is that 
the numerically obtained $\Delta$ increases as $S$ is increased 
in spite of the fact that 
$\Delta$ is regarded as an energy gap in the magnetic field. 
Regarding the $S$ dependence of the energy gap, 
it is well known that 
the Haldane gap of the integer-$S$ Heisenberg chain shows 
exponential decay with respect to $S$\cite{Haldane1,Haldane2}. 
The $S$-dependence of Haldane's expression can be 
compared with reliable numerical estimates 
for spin-$S$ Haldane gaps\cite{comment_Haldane_gap}. 
The present increase in $\Delta$ is contradictory 
to the dependence of the Haldane gap. 
This feature is also different from Eq.~(\ref{Zhitomirsky_delta}). 
Note here that 
the present observation of $\Delta$ in units of $J$ approaching a nonzero limit 
does not contradict the simple expectation that 
the plateau width will vanish in the classical limit 
because the plateau should be measured in comparison with $H_{\rm s}$, 
which is linear in $S$. 
The nonzero limit of $\Delta$ in units of $J$ is different 
from 1/2 in Eq.~(\ref{Zhitomirsky_delta}). 
One cannot, unfortunately, deny the possibility that 
$\Delta$ is overestimated in the present numerical-diagonalization study 
owing to the finite-size effect. 
This is a possible reason for the disagreement. 
Thus, studies tackling calculations of larger clusters 
should be carried out in future.  
Another possible reason for the disagreement is 
that the perturbation treatment around the classical configuration 
may be too rough to properly capture the essential quantum effect 
in the $m=1/3$ state of the kagome-lattice antiferromagnet. 
The roughness may be reduced by additionally taking into account 
the effect from the spin waves; 
an examination along such a direction should be carried out 
in a future study. 


In summary, we have investigated the magnetization process 
of the general spin-$S$ Heisenberg antiferromagnet on the kagome lattice 
by the numerical-diagonalization method. 
We have found that 
a magnetization plateau appears 
at one-third of the height of the saturation 
even in the cases of large $S$. 
Our analysis of the edge fields and the width 
of the plateau suggests that 
the numerical-diagonalization results disagree 
with the equations obtained by real-space perturbation theory. 
The present study, 
based on an {\it unbiased} and {\it non-perturbative theoretical} method, 
presents significant information 
concerning general spin-$S$ antiferromagnets with frustrations, 
which will contribute to future studies. 
In the case of $S=1/2$, the $m=1/3$ state of several frustrated systems 
shows an interesting phase transition between the ferrimagnetic 
state and another state accompanied 
by a novel spin-flop phenomenon\cite{shuriken_lett,HN_kgm_dist,
HNakano_Cairo_lt,Isoda_Cairo_full,HN_TSakai_JJAP_RC,HN_TSakai_kgm_1_3}. 
One of the cases is when the kagome lattice is distorted 
to the $\sqrt{3}\times\sqrt{3}$ type\cite{HN_TSakai_kgm_1_3}. 
Future studies taking 
this distortion into account could help us 
to more clearly estabish the relationship between 
the nine-site structure pointed out in Ref.~\ref{Bergman_JPCM19_145204}
and unbiased numerical data. 
When the kagome lattice is spatially anisotropic, 
a non-Lieb-Mattis-type ferrimagnetic ground state 
of the $S=1/2$ system exists 
near the isotropic point\cite{collapse_ferri2d,Shimokawa_JPSJ}. 
It should be examined in future 
what happens to these nontrivial phenomena when $S\ge 1$. 
Further study on general spin-$S$ systems would greatly contribute 
to our understanding of the frustration effect 
in quantum spin systems. 

\section*{Acknowledgments}
We wish to thank 
Professors 
M.~E.~Zhitomirsky, 
M.~Isoda, 
and 
N. Todoroki 
for fruitful discussions. 
This work was partly supported 
by JSPS KAKENHI Grant Numbers 23340109 and 24540348. 
Some of the computations were 
performed using facilities of 
the Department of Simulation Science, 
National Institute for Fusion Science; 
Center for Computational Materials Science, 
Institute for Materials Research, Tohoku University; 
Supercomputer Center, 
Institute for Solid State Physics, The University of Tokyo;  
and Supercomputing Division, 
Information Technology Center, The University of Tokyo. 
This work was partly supported 
by the Strategic Programs for Innovative Research; 
the Ministry of Education, Culture, Sports, Science 
and Technology of Japan; 
and the Computational Materials Science Initiative, Japan. 
The authors would like to express their sincere thanks 
to the staff of the Center for Computational Materials Science 
of the Institute for Materials Research, Tohoku University, 
for their continuous support 
of the SR16000 supercomputing facilities. 



\begin{thebibliography}{99}
\bibitem{comment_Haldane_gap}
Results for the $S=1$ case were previously obtained 
by density matrix renormalization group (DMRG) 
calculation\cite{White_DMRG}
and by numerical diagonalization\cite{Golinelli_ED}.  
An $S=2$ result was previously obtained 
by DMRG calculation\cite{Shollwoeck_Jolicoeur}; 
an $S=3$ result was previously obtained 
by quantum Monte Carlo calculation\cite{Todo_Kato}.  
Results for the $S=4$ and 5 cases were previously obtained 
by numerical diagonalization\cite{HN_Terai}.
\end{thebibliography}
\end{document}